\documentclass[aps,pre,twocolumn,showpacs,amsmath,amssymb]{revtex4}

\usepackage{graphicx}
\usepackage{dcolumn}
\usepackage{bm}

\begin{document}

\preprint{APS/123-QED}

\title[Ising Ferromagnet with Annealed Vacancies]{A Dilute Ising Ferromagnet on a Hierarchical Lattice with Attractive Biquadratic Interactions}

\author{Daniel P. Snowman}

\affiliation{Department of Physical Sciences, Rhode Island College, \\Providence, Rhode Island 02908}

\date{\today}

\begin{abstract}
  This paper considers a dilute Ising ferromangnet with annealed vacancies and attractive biquadratic interactions.  Phase diagrams have been calculated while varying the temperature and concentration of annealed vacancies in the system while maintaining constant, attractive biquadratic couplings.  These results have been produced using renormalization group analysis with a hierarchical lattice.  Critical exponents have been calculated and each basin of attraction interpreted.
\end{abstract}

\maketitle

\section{Introduction}
The Blume-Emery-Grittiths (BEG) model ~\cite{Blume:PRA71} is a spin-1 Ising model that is useful in probing the nature of systems with fluctuation in magnetization and density.  The Hamiltonian, shown in Equation 1, has the bilinear ($J_{ij}$), biquadratic ($K_{ij}$) and crystal-field interaction ($\Delta_{ij}$) terms as shown in the Hamiltonian in Eq. (1).

\begin{eqnarray}
-\beta H = \sum_{\langle{ij}\rangle} J_{ij} s_i s_j + \sum_{\langle{ij}\rangle} K_{ij} s_i^2 s_j^2 - \sum_{\langle {ij}\rangle}\Delta_{ij}(s_{i}^2 + s_{j}^2)
\nonumber \\
 \text{with} \qquad s_i = 0, \pm 1 \qquad \qquad \qquad \qquad
\end{eqnarray}
 
With these three coupling coefficients we can probe ordering in various systems as we vary temperature $(\sim 1/J)$, concentration $(\sim \Delta/J)$ of annealed vacancies, and the clustering bias $(\sim K/J)$.  

\begin{eqnarray}
 \sum_{\langle{ij}\rangle} H_{ij} (s_i + s_j) + \sum_{\langle{ij}\rangle} L_{ij} (s_i^2 s_j + s_i s_j^2)
\end{eqnarray}

In addition to the bilinear, biquadratic and crystal-field interaction terms we must also consider odd sector contributions to the Hamiltonian, as shown in Eq. (2).  These contributions must be included in order to obtain a complete description of higher order phase transitions arising in our system.

It should be noted that each summation in Eqs. 1 and 2 is over nearest-neighbor $\langle{ij}\rangle$ pairs of our lattice unit structure including the magnetic ($H$) and crystal-field ($\Delta$) interaction terms.  The summation for each of these terms has been shifted from sites to bonds for computational expediency with the overall effect being that $\Delta$ in Eq. (1), and $H$ in Eq. (2), is the chemical potential, and magnetic field, per bond divided by two.  

The introduction of density as an added degree of freedom in Ising systems was originally proposed to investigate the superfluid transition in $He^{3}-He^{4}$ mixtures~\cite{Blume:PRA71}.  Since, these spin-1 Ising systems have been extended and used to explore structural glasses ~\cite{Kirkpatrick:PRB97},  binary alloys, microemulsions ~\cite{Schick:PRB86}, materials with mobile defects, binary fluids, aerogels ~\cite{Maritan:PRL92}, frustrated percolation systems ~\cite{Coniglio:JPF93}, among many others.

Fixed points, critical phenomenon and resulting phase diagrams can be drastically altered due to underlying competing microscopic interactions in various Ising systems.  Mean-field methods ~\cite{Hoston:PRL91, Sellitto:JPF97, Snowman:PhD95, McKay:JAP84} and/or renormalization-group techniques ~\cite{Berker:PRB76, McKay:PRL82, Branco:PRB99, Branco:PRB97, Snowman:JMMM07, Snowman:JMMM08, Snowman:PRE08} have been used in many previous studies to consider a range of qualitatively unique competing interactions using the Blume-Emery-Griffiths model. 

Renormalization-group methods have been employed, in conjunction with hierachical lattices, to investigate the effects of competing biquadratic interactions in a dilute Ising ferromagnet ~\cite{Snowman:PRE08}, competing bilinear interactions in a BEG system ~\cite{Snowman:JMMM08}, competing bilinear interactions ~\cite{McKay:PRL82} in a spin-1/2 Ising model, and simultaneous competition between crystal-field and biquadratic interactions in a BEG ferromagnet ~\cite{Snowman:JMMM07}.  

Other renormalization-group probes yield tricritical and critical end point topologies linking first and second order phase boundaries for the case with $K > 0$.  For negative biquadratic coupling ($K < 0$), mean-field calculations revealed two novel phases: one a high-entropy ferrimagnetic phase and the other displaying antiquadrupolar order, see reference ~\cite{Hoston:PRL91}.

The effects of attractive and repulsive biquadratic interactions on the global phase space was considered by Sellitto et al. ~\cite{Sellitto:JPF97} using the replica symmetric mean-field approximation with quenched disorder in the bilinear interactions.  A spin-glass phase was found with both first and second order transitions from the paramagnetic phase: the order of the transition largely dependent upon the crystal-field interaction. This study also found, for strong repulsive ($K<0$) biquadratic interactions, an antiquadrupolar phase and at lower temperatures, an antiquadrupolar spin-glass phase.

Kabakcioglu et al. considered the effects of quenched random fields ~\cite{ Kabak:PRL99} and Falicov et al. the effects of quenched random bonds ~\cite{Falicov:PRL96} upon ordering, criticality and resulting phase diagrams.  Random crystal fields were the focus of Branco et. al using real-space RG ~\cite{Branco:PRB99,Branco:PRB97} and mean-field approximations for both Blume-Capel and Blume-Emery-Griffiths model Hamiltonians, respectively.

The current study complements these earlier works as it considers a dilute Ising ferromangnet with annealed vacancies and attractive biquadratic interactions.  The effects of temperature and concentration of annealed vacancies upon ordering for a series of constant, attractive biquadratic couplings have been probed via calculation of several phase diagrams.  Each phase diagram produced using renormalization group analysis with a hierarchical lattice.  Critical exponents have been calculated for alll higher-order transitions and each basin of attraction interpreted.

\section{Hierarchical Lattices and Renormalization Group}
The basic recipe for generating an infinite hierarchical lattice from its basic unit is to repeatedly replace each nearest-neighbor interaction by the basic unit itself.  The construction of a generic hierarchical lattice is shown in Figure 1.  Figure 2 illustrates the construction of the hierarchical lattice ~\cite{Berker:JPC79,Kauffman:PRB81} used for this study, with the difference being the complexity of the basic generating unit. 

Since hierarchical lattices yield exact renormalization group recursion relations for the coupling coefficients, phase diagrams can be produced and critical exponents determined very accurately.  Therefore the results presented in this study may be considered exact on the rather specialized lattice, or, they may be used as approximations into these systems modeled on more realistic lattices.  A range of very difficult problems has been subjected to study using hierarchical lattices as the medium.   For example, random-bond~\cite{ Andelman:PRB84}, random-field ~\cite{Falicov:PRB95}, spin glass ~\cite{Snowman:JMMM08, Migliorini:PRB98}, frustrated ~\cite{McKay:PRL82, Snowman:JMMM07,Snowman:PRE08}, directed-path ~\cite{daSilv:PRL04} and dynamic scaling ~\cite{Stinchcombe:JPA86} systems have all been investigated and better understood using these specialized lattices.

The renormalization-group solution for a hierarchical model, such as Figures 1 and 2, essentially reverses the construction process.  Internal degrees of freedom are eliminated with each renormalization by summing over all configurations of the internal sites (represented by solid black dots in Figures 2a and 2b, and by {$\sigma_i, \sigma_j$} in Equation 6).

\begin{figure}
\begin{center}
\leavevmode
\includegraphics[scale=0.5]{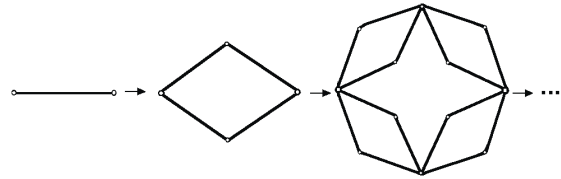} 
\end{center}
\caption{The construction of an infinite hierarchical lattice by repeatedly replacing each nearest-neighbor interaction by the basic unit itself (Berker and Ostlund ~\cite{Berker:JPC79}).}
\label{fig:Figure 1}
\end{figure}

\begin{figure}
\begin{center}
\leavevmode
\includegraphics[scale=0.5]{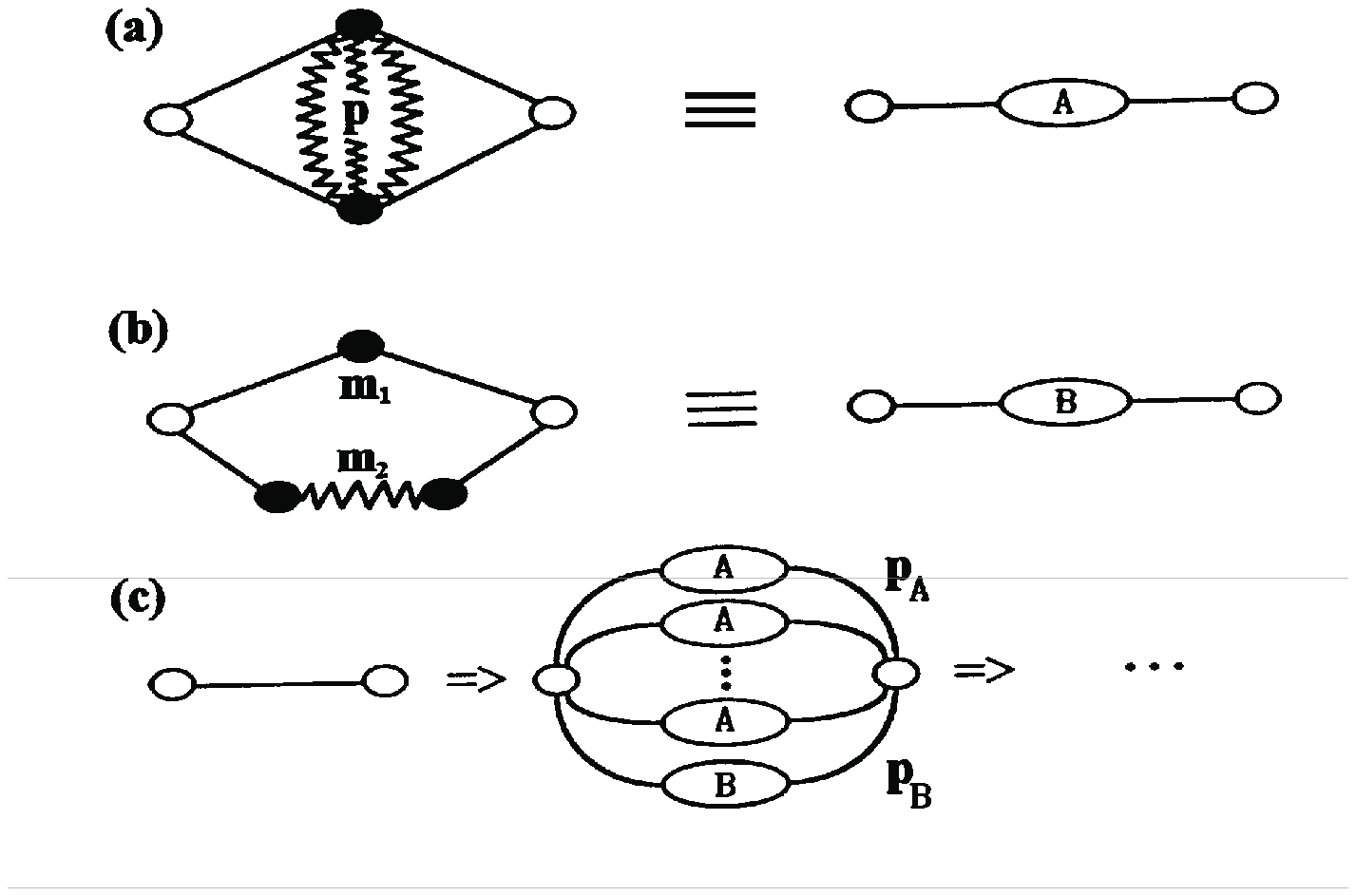} 
\end{center}
\caption{Construction of the hierarchical lattice used in this investigation. (Reprinted from Journal of Magnetism and Magnetic Materials, 314, 69-74 (2007) D. P. Snowman,  with permission from Elsevier)}
\label{fig:Figure 2}
\end{figure}

With each renormalization, or rescaling, in our system we demand the partition function remain unchanged.  This allows us to derive recursion relations that relate the exchange interactions at the two length scales.  The new effective interactions $J'$, $K'$, and $\Delta'$ are separated by a distance $l'$ which is $b$ lattice constants in the original system, where $b$ is the length rescaling factor of the renormalization-group transformation.

\begin{eqnarray}
\zeta_{l'}(J',K',\Delta ')= \zeta_{l}(J,K,\Delta) \\
\nonumber\\
\text{with} \; l' = b l \qquad \qquad 
\end{eqnarray}

\begin{eqnarray}
\zeta_l = \sum_{s_i, s_j} \exp[-\beta H] = \sum_{s_i, s_j} R_l (s_i,s_j)\\
\text{with}\; R_l(s_i,s_j) = \sum_{\sigma_i,\sigma_j}\exp[-\beta H]
\end{eqnarray}

\begin{eqnarray}
\zeta_{l'} = \sum_{s_i^{'},s_j^{'}} \exp[-\beta H'] = \sum_{s_i^{'}, s_j^{'}} R_l (s_i^{'},s_j^{'}) \\
\text{with} \; R_l(s_i^{'},s_j^{'}) = \sum_{s_i^{'}, s_j^{'}}\exp[J's_i s_j \nonumber \\ \; + K' s_i^2 s_j^2  - \Delta ' (s_i^2+s_j^2) + \widetilde{G}']
\end{eqnarray}
  
where $\widetilde{G}'$ is a constant used to calculate the free energy. 

Equating the contributions to the partition function, $R_{l}(s_i,s_j)$ and $R_{l^'}(s_i,s_j)$ - corresponding to the same fixed configuration of end spins, ${s_i, s_j}$ and ${s_i^{'}, s_j^{'}}$ - at the two different length scales, $l$ and $l^'$, allows us to calculate the actual renormalzation-group transformations.  Thus relating the interaction strengths at the two length scales, $l$ and $l^'$, can be derived: $J' (J, K, \Delta), K' (J, K, \Delta),$ and $\Delta ' (J, K, \Delta)$.  The derivation of these relations is presented in more detail in Section 5.

Parameter space is explored, phase diagrams mapped and transitions characterized using these recursion relations in conjunction with the initial values of $J, K$ and $\Delta$.  Repeated iteration of the recursion relations from an initial starting point carries the system along a renormalization-group trajectories and eventually to a sink.

\begin{eqnarray}  
J' = R_J (J, K, \Delta)	\\
\nonumber\\
K' = R_K (J, K, \Delta)	\\	
\nonumber\\  			 
\Delta ' = R_{\Delta} (J, K, \Delta) \\
\nonumber
\end{eqnarray}

Two types of components (see Fig. 2a and b) have been included in the basic unit, similar to ~\cite{McKay:PRL82, McKay:JAP82,Snowman:JMMM07,Snowman:JMMM08,Snowman:PRE08}, for the calculations presented in this paper.  One type of component, type A, allows internal spins to interact via a cross-link feature.  The strength of the cross-link interaction is varied via the parameter p, with the case of $p=0$ yielding the hierarchical model equivalent ~\cite{Berker:JPC79} to the Migdal-Kadanoff ~\cite{Migdal75, Kadanoff76} decimation-bond moving scheme in two dimensions.    

A second type of component, type B, allows the end spins to interact via two different length connecting paths.  As shown in Figure 2b, the first path has $m_1$ pairs of spins and the second $m_2$ pairs of spins.  The overall connectivity of the system is varied using two parameters, $p_A$ and $p_B$.  These parameters, represent the number of unit structures, either type A or type B (see Fig. 2), used to form the basic generating unit for the hierarchical lattice as shown in Fig. 2c.  The connectivity parameters, $(p, m_1, m_2, p_A, p_B) = (4, 8, 9, 40, 1)$, used in the present investigation parallel those used in previous studies, see references ~\cite{McKay:PRL82, McKay:JAP82,Snowman:JMMM07,Snowman:JMMM08,Snowman:PRE08}, probing the effects of various types of competing interactions on a similar lattice.  

\begin{eqnarray}
J^{*} = R_J(J^{*}, K^{*}, \Delta^{*}) \\
\nonumber\\
K^{*} = R_K(J^{*}, K^{*}, \Delta^{*})\\
\nonumber\\				
\Delta^{*} = R_{\Delta}(J^{*}, K^{*}, \Delta^{*})\\
\nonumber
\end{eqnarray}

\section{Phase Transition Characterization}
Numerically differentiating the free energy density with respect to the appropriate variables yields values for the magnetizations, densities and nearest neighbor correlations for our system.  The free energy density (dimensionless free energy per bond ), $f$, can be expressed as

\begin{equation}
f = {- \frac{\beta F}{N_{b}}} = \sum_{n=1}^{\infty}b^{-nd}{G'^{(n)}}(J^{(n-1)},K^{(n-1)},\Delta^{(n-1)}  
\end{equation}

where $F$ is the Helmhotz free energy and $N_b$ denotes the total number of bonds in the system.  The free energy density consists of a sum, over all iterations of the renormalization-group transformation, of the contributions $G'^{(n)}$ to the free energy density due to the degrees of freedom removed during each transformation.  With each rescaling, or renormalization, the length scale of the system is reduced by a factor of $b$ and the number of spins by a factor of $b^d$.

	The magnetization, $m \equiv \frac{M}{N_s} = \frac{N_b}{N_s}\frac{\delta f}{\delta H}$, can be calculated by numerically measuring the shift in the free energy density with a small shift in the magnetic field, where $N_s$ is the number of sites.  Using a similar approach yields the density, $\rho \equiv \frac{N_b}{N_s}\frac{\delta f}{\delta \Delta}$ .  Nearest neighbor correlations of the bilinear, $\langle s_i s_j \rangle = \frac{N_b}{N_s}\frac{\delta f}{\delta J}$, and biquadratic, $\langle s_i^2 s_j^2 \rangle = \frac{N_b}{N_s}\frac{\delta f}{\delta K}$, exchange interactions also aid in interpreting resulting phases and transitions.  

Using the four thermodynamic quantities discussed above, transitions between the various basins of attraction are characterized.  First order transitions reveal themselves by discontinuities in densities, magnetizations, or other first derivatives of the free energy.  Critical transitions exhibit no such discontinuities in any of the order parameters.  Exact expressions for the recursion relations allow us to calculate critical scaling exponents for our system.  The next section explores this in greater detail.

\section{Recursion Relations}
For each fixed end spin configuration, equating the contributions to the partition function, $R_{l}(s_i,s_j)$ and $R_{l^'}(s_i,s_j)$, from the two length scales - as introduced in section 2 -  allows us to write the following equalities for the type A structure shown in Figure 2a.  

\begin{widetext}
\begin{eqnarray}
R_l[1,1] & = & \exp[-4 \Delta]+2 \exp[-2 J+2 K-\Delta (6+p)]
\nonumber \\
& & + 2 \exp[2 J+2 K-\Delta (6+p)]+
  \exp[J (-4+p)+K (4+p)-\Delta (8+2 p)]
\nonumber \\
& & + 2 \exp[-J p+K (4+p)-\Delta (8+2 p)]+
  \exp[J (4+p)+K (4+p)-\Delta (8+2 p)] 
\nonumber \\
& & = \exp[J' + K' - 2 \Delta' + \tilde{G}] = R_{l'}[1,1],
\end{eqnarray}

\begin{eqnarray}
R_l[1,0] & = & \exp[-2 \Delta]+2 \exp[-J+K-\Delta (4+p)]+2 \exp[J+K-\Delta (4+p)]
\nonumber \\
& & + \exp[J (-2+p)+K (2+p)-\Delta (6+2 p)]+2 \exp[-J p+K (2+p)-\Delta (6+2 p)]
\nonumber \\
& & +\exp[J (2+p)+K (2+p)-\Delta (6+2 p)] = \exp[-\Delta' + \tilde{G}] = R_{l'}[1,0],
\end{eqnarray}

\begin{eqnarray}
R_l[1,-1] & = & \exp[-4 \Delta]+4 exp[2 K-\Delta (6+p)]
\nonumber \\
& & + 2 \exp[-J p+K (4+p)-Delta (8+2 p)]+2 
    \exp[J p+K (4+p)-Delta (8+2 p)]
\nonumber \\
& & = \exp[- J' + K' - 2\Delta' + \tilde{G}] = R_{l'}[1,-1],
\end{eqnarray}

\begin{eqnarray}
R_l[0,0] & = & 1 + 4 \exp[-\Delta (2+p)]+2 \exp[-J p+K p-\Delta (4+2 p)]
\nonumber \\
& & + 2 \exp[J p+K p-\Delta (4+2 p)] = \exp[\tilde{G}] = R_{l'}[0,0],
\end{eqnarray}
\end{widetext}

Eqs. 16-19 can now be manipulated algebraically to yield expressions that relate the coupling coefficients between the two length scales for the type A unit structure.

\begin{eqnarray}
J_A^{'} = \frac{1}{2}\log{\frac{R_{l'}(1,1)}{R_{l'}(1,-1)}}\qquad\qquad\qquad\\
\nonumber\\
K_A^{'} = \frac{1}{2}\log{\frac{R_{l'}(1,1) R_{l'}(1,-1) R_{l'}^2(0,0)}{R_{l'}^4(1,0)}}\\
\nonumber\\
\Delta_A^{'} = \log{\frac{R_{l'}(0,0)}{R_{l'}(1,0)}}\qquad\qquad\qquad\\
\nonumber\\
\widetilde{G}_A^{'} = \log{R_{l'}(0,0)}\qquad\qquad\qquad\\
\nonumber
\end{eqnarray}

The type B unit structure has recursion relations with the same form as in Eqs. 20-23, but the expressions (Eqs.16-19) for the underlying $Rl(s_i,s_j)$ differ significantly.  Combining the recursion relationships for both types of structures (type A and type B as shown in Fig. 2), the renormalization relationships become

\begin{eqnarray}
J^{'} = p_A J_A^{'} + p_B J_B^{'}\\
\nonumber\\
K^{'} = p_A K_A^{'} + p_B K_B^{'}\\
\nonumber\\
\Delta{'} = p_A \Delta_A^{'} + p_B \Delta_B^{'}\\
\nonumber
\end{eqnarray}

Critical exponents can be extracted from the exact recursion relations above via linearization in the vicinity of the second-order transition under investigation.  That is, 

\begin{eqnarray}
J^{'}-J^{*} & = & T_{JJ} (J-J^{*})+T_{JK} (K-K^{*})\nonumber\\
& & + T_{J{\Delta}}(\Delta-\Delta^{*}),
\end{eqnarray}

\begin{eqnarray}
K^{'}-K^{*} & = & T_{KJ} (J-J^{*})+T_{KK} (K-K^{*})\nonumber\\
& & + T_{K{\Delta}}(\Delta-\Delta^{*}),
\end{eqnarray}

\begin{eqnarray}
\Delta^{'}-\Delta^{*} & = & T_{{\Delta}J}(J-J^*) + T_{{\Delta}K}(K-K^*)\nonumber\\
& & + T_{{\Delta}{\Delta}}(\Delta-\Delta^{*}),
\end{eqnarray}

where  $T_{JJ} = \frac{\delta J'}{\delta J}$, $T_{KJ} = \frac{\delta K'}{\delta J}$, etc. and are evaluated at the fixed point in question.  The critical relations above can be represented as a recursion matrix, with elements $T_{XY}$ and eigenvalues of the form

\begin{equation}
\Lambda_l = b^{y_l} \qquad \text{where} \qquad l=2,4,6
\end{equation}

where $b$ is the length rescaling factor (in our case $b=2$) and $y_l$ represents the corresponding critical exponent for the $l^{th}$ eigenvalue.  A similar linearization for the odd sector contributions ($H$ and $L$) yields critical scaling exponents $y_1$ and $y_3$. 

\section{Results}
The results of our investigation, detailed below, probed the effects upon ordering of varying the temperature ($1/J$) and vacancy concentration ($~{\Delta}/J$) for a series of planes of constant biquadratic coupling $K/J$.  In each plane in parameter space, bulk phases have been mapped out by exhaustively analyzing the nature of renormalization-group trajectories that arise for various initial starting parameters ($J,K, \Delta$).  This exploration reveals three qualitatively unique regions (or basins of attraction), each sharing renormalization-group trajectories that flow to common sinks, as detailed in Table I.

\begin{table}
\begin{tabular}{|l|l|l|}
\hline  Phase & Sink & Characteristics \\ 
\hline  Paramagnetic I & $J \rightarrow 0$ & Low Concentration \\ 
  &  $K \rightarrow 0$ & Nonmagnetic impurities\\ 
  &  $\Delta \rightarrow -\infty$ & Dilute Sublattice II\\ 
\hline  Ferromagnetic I& $J \rightarrow +\infty$ & Mag. Order \\ 
  & $K \rightarrow -\infty$ &  \\ 
  & $\Delta \rightarrow -\infty$ & \\ 
\hline  Paramagnetic II& $J \rightarrow 0$ & High Concentration \\ 
  &  $K \rightarrow 0$ & Nonmagnetic impurities \\ 
  &  $\Delta \rightarrow +\infty$&  \\ 
\hline  Ferromagnetic II& $J \rightarrow +\infty$ & Mag. Order \\ 
  & $K \rightarrow -\infty$ &  \\ 
  & $\Delta \rightarrow +\infty$ & \\
\hline 
\end{tabular}
\vspace{3mm}
\caption{Phases and Corresponding Sinks}
\end{table}

The two paramagnetic phases arising, Dense Paramagnetic and Dilute Paramagnetic, are distinguished from one another via the flow of the crystal-field interaction term.  A flow to -$\infty$ corresponding to a high density of occupied sites, whereas a flow to +$\infty$ corresponds to a system dominated by nonmagnetic spin sites and much too dilute for magnetic ordering to occur.  A Dense Ferromagnetic, complete with crystal-field interaction flowing to -$\infty$, also arise in phase space, with a renormalization group flow that sees the bilinear ($J$) interaction flowing to +$\infty$, and, the biquadratic ($K$) coupling flowing to -$\infty$ .  The ferromagnetic phase can be reached by decreasing the temperature, or, by decreasing the concentration of nonmagnetic impurities.

\begin{figure}
\begin{center}
\leavevmode
\includegraphics[scale=0.5]{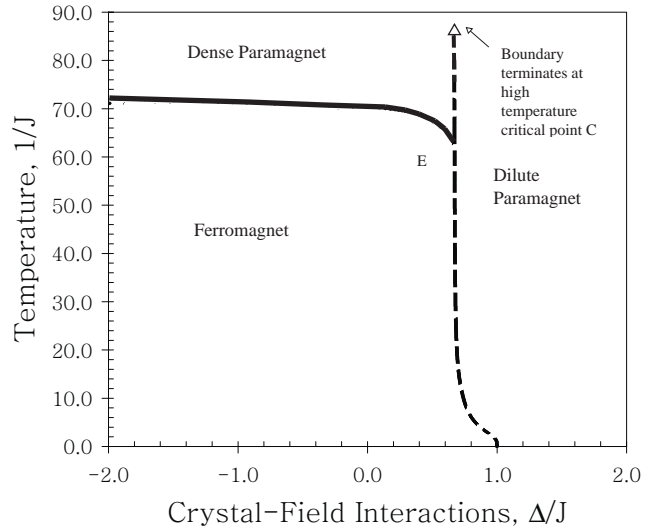} 
\end{center}
\caption{Parameter space, with K/J = 1 and 
($p, m_1, m_2, p_A, p_B$) = (4, 8, 9, 40, 1), depicting different basins of attraction and associated phases with critical endpoint (E) and critical point (C). Solid lines represent second-order transitions, whereas dashed lines represent first-order transitions.}
\label{fig:Figure 3}
\end{figure}

In the plane of constant biquadrupolar coupling $K/J = 1$ we find each of the phases detailed above.  The Dilute Paramagnetic phase exists at all temperatures for larger, positive values of the crystal-field interaction (${\Delta}/J$); the lack of magnetic order consistent with the expectations for a lattice dominated by nonmagnetic impurities.  At high temperatures ($1/J > 70$) we find the Dilute Paramagnetic phase separated from the Dense Paramagnetic phase via a first order phase boundary.  This phase boundary is entirely vertical thus it is traversed only via a changing crystal-field interaction.  The first order boundary separating the two paramagnetic phases terminates at a high temperature critical point C.

\begin{figure}
\begin{center}
\leavevmode
\includegraphics[scale=0.5]{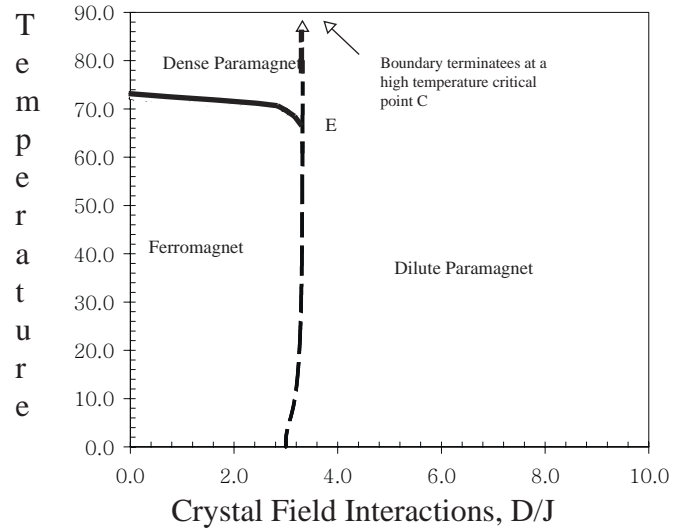} 
\end{center}
\caption{Parameter space, with K/J = 5 and 
($p, m_1, m_2, p_A, p_B$) = (4, 8, 9, 40, 1), depicting different basins of attraction and associated phases with critical endpoint (E) and critical point (C).  Solid lines represent second-order transitions, whereas dashed lines represent first-order transitions.}
\label{fig:Figure 4}
\end{figure}

In this same plane, $K/J = 1$, the Dense Paramagnetic phase is driven to magnetically order to a Ferromagnetic phase with a decrease in temperature via second order transition at approximately $1/J$ $\sim$ $70$.  This line of criticality terminates upon intersection with the line of first order transitions, at critical endpoint E.  The Ferromagnetic phase persists for the more negative values of the crystal-field interaction, corresponding to a low concentration of nonmagnetic impurities on the lattice allowing magnetic order to propagate.  An increase in the crystal field drives the system to disorder to the Dilute Paramagnetic phase via a first order transition.  This line of first order transition curves at low temperatures allowing for the possibility of the Dilute Paramagnetic phase to magnetically order with a decrease in temperature for a fixed concentration of occupied sites, e.g. ${\Delta}/J = 0.8$.

An increase of the biquadratic coupling to $K/J = 5$ results in a shift of the first order line of transitions to larger ${\Delta}/J$.  This first order phase boundary separates the two paramagnetic states at high temperatures, and the ferromagnetic and dilute paramagnetic phases at intermediate and low temperatures.  The low temperarture phase boundary no longer has the curved feature allowing for the ordering of the Dilute  Paramagnetic phase with a decrease in temperature, as seen in the $K/J = 1$ plane.  Transition from the Ferromagnetic phase to the Dilute Paramagnetic phase driven only with a change in the concentration of the nonmagnetic impurities.

\begin{figure}
\begin{center}
\leavevmode
\includegraphics[scale=0.5]{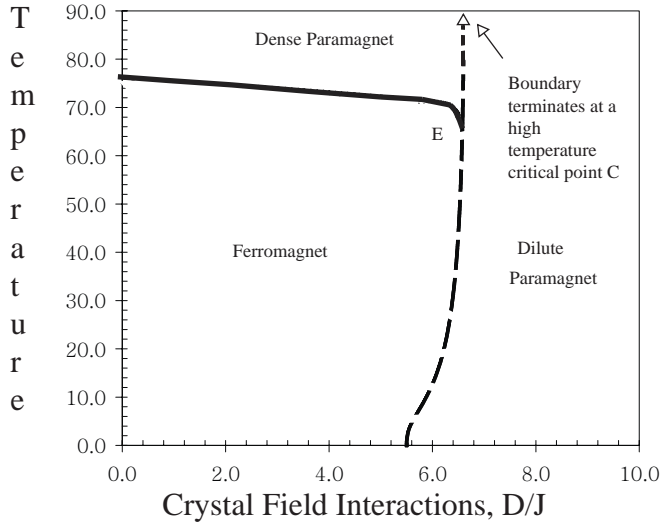} 
\end{center}
\caption{Parameter space, with K/J = 10 and 
($p, m_1, m_2, p_A, p_B$) = (4, 8, 9, 40, 1), depicting different basins of attraction and associated phases with critical endpoint (E).  Solid lines represent second-order transitions, whereas dashed lines represent first-order transitions.}
\label{fig:Figure 5}
\end{figure}

Further increase of the biquadratic coupling to $K/J = 10$ reveals a phase diagram that is qualitatively the same complete with critical point (C) and critical endpoint (E) topologies.  However, at low temperatures the boundary separating the Ferromagnetic and Dilute Paramagnetic phases has curved in the opposite direction as originally observed (in the $K/J = 1$ plane).  This phase diagram revealing the possibility, at certain concentrations (e.g. ${\Delta}/J = 5.6)$, that the Ferromagnetic phase can surprisingly disorder to the Dilute Paramagnetic phase with a decrease in temperature.

\begin{figure}
\begin{center}
\leavevmode
\includegraphics[scale=0.5]{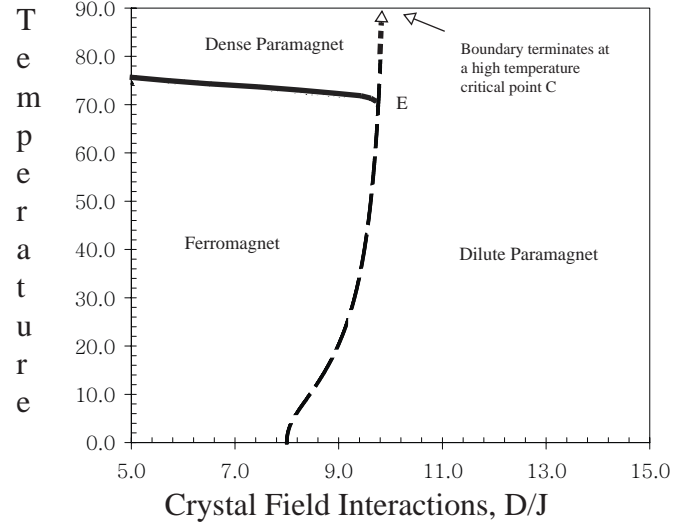} 
\end{center}
\caption{Parameter space, with K/J = 15 and 
($p, m_1, m_2, p_A, p_B$) = (4, 8, 9, 40, 1), depicting different basins of attraction and associated phases with critical endpoint (E) and critical point (C). Solid lines represent second-order transitions, whereas dashed lines represent first-order transitions.}
\label{fig:Figure 6}
\end{figure}

In the last two phase diagrams considered, $K/J = 15$ and $K/J = 20$,  additional increase in the clustering  bias (or biquadratic coupling) drives the line of first order transitions to larger crystal-fields (${\Delta}/J$) with an increasing curvature at low temperatures.  Interestingly, in both of these planes, it is possible to magnetically order the Dense Paramagnetic phase via a decrease in temperature into the Ferromagnetic phase.  And, with an additional decrease in temperature, at certain concentrations (e.g. ${\Delta}/J = 11.0$ in the $K/J = 20$ plane), it is possible for the Ferromagnetic phase to disorder to the Dilute Paramagnetic phase.  An interesting reentrant phenomena albeit to two different paramagnetic states.

\begin{figure}
\begin{center}
\leavevmode
\includegraphics[scale=0.5]{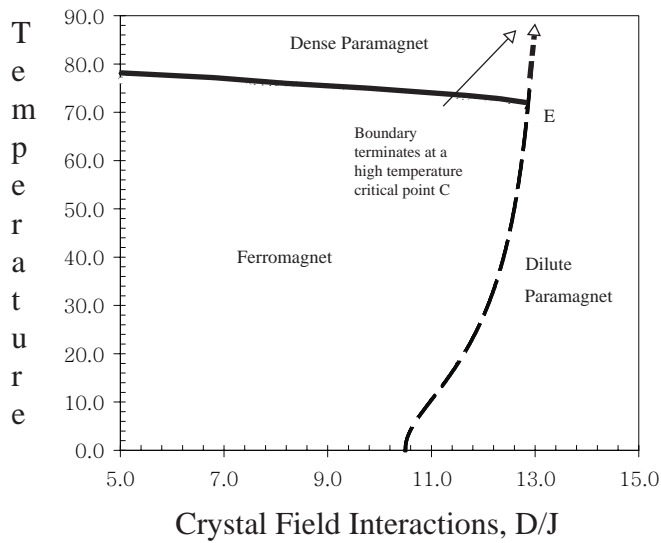} 
\end{center}
\caption{Parameter space, with K/J = 20 and 
($p, m_1, m_2, p_A, p_B$) = (4, 8, 9, 40, 1), depicting different basins of attraction and associated phases with critical endpoint (E) and critical point (C). Solid lines represent second-order transitions, whereas dashed lines represent first-order transitions.}
\label{fig:Figure 7}
\end{figure}

In each plane (of constant biquadratic coupling) considered in the present investigation, the high temperature transition from the ferromagnetic phase to the dense paramagnetic phase was found to be second-order.  This critical line has been probed and critical exponents calculated by linearizing the recursion relations, as discussed in Section 4, while maintaining five scaling fields associated with $J, K, \Delta, H$, and $L$.  Associated with the scaling fields $J, K,$ and $\Delta$ calculation of the eigenvalues of the resulting recursion matrix yields:  two relevant eigenvalues, $\Lambda_2 = 19.6$ and $\Lambda_4 = 2.00$, corresponding to critical scaling exponents of $y_2=4.29$ and $y_4=1.00$, respectively; and, an irrelevant eigenvalues with $\Lambda_6 = -1.63$.  A similar analysis associated with scaling fields $H$ and $L$ yields two relevant eigenvalues: 
$\Lambda_1 = 8.00$ and $\Lambda_3 = 2.00$, corresponding to critical scaling exponents of $y_2=3.00$ and $y_4=1.00$.

\section{Summary}
This paper considers a dilute Ising ferromangnet with annealed vacancies and attractive biquadratic interactions.  Phase diagrams have been calculated while varying the temperature and concentration of annealed vacancies in the system while maintaining constant, attractive biquadratic couplings.  These results have been produced using renormalization group analysis with a hierarchical lattice.  Critical exponents have been calculated and each basin of attraction interpreted.

This study has considered several planes of constant, attractive biquadratic couplings.  Three basins of attraction were found each corresponding to a unique bulk phase in parameter space.  Two paramagnetic regions (Dense Paramagnetic and Dilute Paramagnetic), and a Ferromagnetic region.  First-order boundaries separate the Dense Paramagnetic/Dilute Paramagnetic regions at high temperatures; this boundary terminating at a high temperature critical point C.  The Ferromagnet region is separated from the Dilute Paramagnetic region also with a first-order line of transitions.  The high temperature Dense Paramagnetic/Ferromagnetic second-order line of transitions terminates at a critical endpoint E.  An increase in the biquadratic coupling results in similar phase diagrams with greater curvature introduced in the low temperature regime thus allowing transitions between the Ferromagnetic and Dilute Paramagnetic regions with a change in crystal-field interaction and/or temperature.  The results presented in this paper will provide further insight into the role of nonmagnetic impurities upon ordering transitions in ferromagnetic materials as we vary the density and clustering bias in the system.

\section{acknowledgements}
The author would like to thank Rhode Island College for research release time, and institutional resources, in support of this work.


\end{document}